**Tailoring Interlayer Exchange Coupling of Ferromagnetic Films Across MgO with Fe Nanoclusters**

Jared J. I. Wong, Luciana Ramirez, A. G. Swartz, A. Hoff, Wei Han, Yan Li, and R.K. Kawakami[‡]

Department of Physics and Astronomy, University of California, Riverside, CA 92521

[‡]e-mail: roland.kawakami@ucr.edu

## ABSTRACT

We investigate the interlayer exchange coupling in Fe/MgO/Fe and Fe/MgO/Co systems with magnetic Fe nanoclusters embedded in the MgO spacer. Samples are grown by molecular beam epitaxy (MBE) and utilize wedged MgO films to independently vary the film thickness and the position of the Fe nanoclusters. Depending on the position of the Fe nanoclusters, the bilinear coupling ($J_1$) exhibits strong variations in magnitude and can even switch between antiferromagnetic and ferromagnetic. This effect is explained by the magnetic coupling between the ferromagnetic films and the magnetic nanoclusters. Interestingly, the coupling of Fe nanoclusters to a Co film is 160% stronger than their coupling to a Fe film (at MgO spacing of 0.56 nm). This is much greater than the coupling difference of 20% observed in the analogous thin film systems (i.e. Fe/MgO/Co vs. Fe/MgO/Fe), identifying an interesting nano-scaling effect related to the coupling between films and nanoclusters.

# I. INTRODUCTION

The scaling of magnetic materials down to nanoclusters has led to novel magnetic and spin dependent properties.[1-9] One of the most fascinating magnetic properties is the interlayer exchange coupling (IEC) across MgO, which originates from spin-dependent tunneling between the ferromagnetic layers.[10-14] An interesting issue is the effect of nanoscaling on the behavior of IEC across MgO. Recently, theoretical studies have explored some aspects of this issue and predict that the IEC can be strongly affected by the type and position of impurities in the MgO.[15, 16] Experimentally, however, the role of nanoclusters or other impurities on the IEC across MgO remains an open question.

In this study, we utilize the magneto-optic Kerr effect (MOKE) to examine the IEC in Fe/MgO/Fe and Fe/MgO/Co systems with magnetic Fe nanoclusters (NC) embedded in the MgO spacer. Samples are grown by molecular beam epitaxy (MBE) and utilize wedged MgO films to independently vary the film thickness and the position of the Fe NC. By varying the position of the Fe NC within the MgO spacer, the bilinear coupling ($J_1$) exhibits strong variations in magnitude and can even switch between antiferromagnetic and ferromagnetic. We find that the main features of the data are explained by a model that assumes only pair-wise coupling. Surprisingly, the IEC between Fe NC and a FM film exhibits a strong dependence on the film material (Co vs. Fe): the Fe NC-Co layer coupling is 160% stronger than the Fe NC-Fe layer coupling. When compared to the analogous thin-film systems at comparable spacing, the coupling in Fe/MgO/Co is only 20% stronger than the coupling in Fe/MgO/Fe, showing there is an enhanced material dependence of the IEC due to nano-scaling effects.

# II. EXPERIMENTAL PROCEDURES

## A. Sample growth

All samples are grown on double-side-polished MgO(001) substrates using molecular beam epitaxy (MBE) in ultra high vacuum (UHV) with a base pressure of ~$1\times10^{-10}$ torr. The MgO material is deposited by electron beam evaporation at a rate of ~0.2 nm/min. The other materials (Co, Fe and Ag) are deposited from thermal effusion cells at a rate of ~0.15 nm/min. Deposition rates are determined by a quartz deposition monitor and are verified through reflection high energy electron diffraction (RHEED) intensity oscillations. Substrates are prepared by a pre-rinse in DI water and then annealed at 600 °C in UHV until a clear RHEED pattern is achieved (~ 45 min.). The substrate is subsequently cooled to 350°C followed by the deposition of a 10 nm MgO buffer layer which produces sharp, streaky RHEED patterns as shown in Figure 1(C) (taken at room temperature).[10, 17] Besides improving the surface quality, the buffer layer also helps eliminate any contamination that may arise from impurities in the substrate.[18]

Two types of samples are investigated in this study [Fig. 1(A) and 1(B)]. Both have a "free" magnetic layer with low coercivity (~30 Oe), a "hard" magnetic layer with high coercivity (~350 Oe), and an MgO spacer layer which may have embedded magnetic NC. For the "Fe/MgO/Fe" samples [Fig. 1(A)], the free layer consists of a 15 nm Fe layer grown on top of the MgO buffer layer at RT and annealed at 450°C for 15 minutes, leading to a sharp RHEED pattern [Fig. 1(D)]. For the "Fe/MgO/Co" samples [Fig. 1(B)], an additional 4 monolayers (ML) of Co is deposited at RT on top of the Fe to complete the free layer. Typical Co deposition on Fe exhibits RHEED oscillations and a sharp RHEED pattern [Fig. 1(E) and 1(F)], confirming the epitaxial growth with bcc structure.[19-21]

The MgO spacer is deposited at RT and wedged films of various geometries are used to vary the MgO thickness and/or the position of embedded magnetic NC within the MgO. The magnetic NC consist of ¼ ML of Fe deposited at RT. It is well known that Fe grows as nanoclusters on top of MgO.[22-25] After completing the MgO spacer, a hard layer consisting of Co(50 nm)/Fe(5 nm) and a capping layer consisting of MgO(10nm)/Ag(10nm) are deposited at RT.

### B. Magneto-optic Kerr Effect (MOKE) Measurement

Magnetic characterization of the sample is done by *ex-situ* longitudinal magneto-optic Kerr Effect (MOKE) measurement with the applied magnetic field along the [100] in-plane direction of the Fe. The laser beam is incident through the MgO substrate to measure both the free and hard layer magnetizations. A typical hysteresis loop [Figure 1(G), dashed curve] exhibits a switching of the free layer (~30 Oe) followed by a switching of the hard layer (~350 Oe). Minor hysteresis loops [Figure 1(G), solid curve] are measured to determine $J_1$ according to $J_1 = H_1 M_{free} t_{free}$, where $H_1$ is center position of the minor loop, $M_{free}$ is the magnetization of the free layer (black arrow), and $t_{free}$ is the free layer thickness.[10-12] A negative $H_1$ indicates antiferromagnetic (AF) coupling ($J_1 < 0$) and a positive $H_1$ indicates FM coupling ($J_1 > 0$).  For some cases of low MgO thickness, the AF coupling becomes so strong that the hard layer does not remain pinned and this method cannot be used to determine $J_1$.[11]

### III. RESULTS

#### A. Interlayer exchange coupling without nanoclusters

We first investigate $J_1$ as a function of MgO thickness in both the Fe/MgO/Fe and Fe/MgO/Co systems by using the MgO wedge structure shown in Fig. 2(A).  To avoid sample-to-sample variations, the Fe/MgO/Fe and Fe/MgO/Co systems are grown on the same sample by depositing the 4 ML Co layer on half of the sample. This sample, denoted as Sample A, allows us to directly compare couplings found in Fe/MgO/Fe and Fe/MgO/Co and investigates any material dependence in IEC.

Figure 2(B) shows the detailed dependence of $J_1$ on MgO thickness for Fe/MgO/Fe (white squares) and Fe/MgO/Co (black circles) obtained by scanning MOKE along the MgO wedge. At high MgO thicknesses (>0.85 nm), both systems show very little to no coupling (below our measurement resolution of ~0.005 erg/cm$^2$).  As the MgO thickness decreases below ~0.85 nm, the coupling is AF and increases in strength with decreasing MgO thickness.  The curves for Fe/MgO/Fe and Fe/MgO/Co are similar for MgO thickness down to ~0.65 nm.  Below MgO thickness of 0.65 nm, the curves deviate from each other with maximum measured $J_1$ = -0.54 erg/cm$^2$ for Fe/MgO/Fe and $J_1$ = -0.70 erg/cm$^2$ for Fe/MgO/Co at MgO thickness of 0.47 nm (~ 30% difference).

In the region of MgO thickness below 0.47 nm (~2.1 ML) the coupling cannot be determined through minor loop analysis due to strong AF coupling.  Qualitatively, in this low MgO thickness region, the coupling for Fe/MgO/Co changes very drastically to FM coupling at an MgO thickness of ~0.43 nm.  For Fe/MgO/Fe, the coupling is strongly AF down to MgO thickness of ~ 0.31 nm and the coupling transitions to FM coupling at an MgO thickness of ~0.27 nm.

#### B. Effect of Fe nanoclusters on the interlayer exchange coupling

We explore the effect of embedding Fe NC within the MgO spacer in both the Fe/MgO/Fe and Fe/MgO/Co systems.  To systematically study the dependence of $J_1$ on both NC position and MgO thickness, we use the MgO double-wedge spacer shown in Fig. 3. The two MgO wedges are grown along perpendicular directions and have the Fe NC sandwiched in between. By scanning along the double wedge from B to D, we can obtain $J_1$ as a function of MgO thickness while keeping the NC at the same relative position (i.e. in the middle of the spacer). By scanning from A to C, we are able to determine $J_1$ as a function of NC position while keeping the total MgO thickness fixed.[26, 27] At point A the NC is located at the hard layer/MgO interface, and at point C the NC is located at the free layer/MgO interface. For this study, we focus primarily on the line scans parallel to A-C, to systematically measure $J_1$ versus NC position at various total thicknesses of MgO.

First, we examine the coupling in the Fe/MgO/Fe system with NC, which we denote as Sample B. The line cuts of $J_1$ versus NC position are shown in Fig. 4(A) at MgO thickness of 1.04 nm (blue circles), 0.85 nm (green diamonds), 0.75 nm (orange triangles), 0.70 nm (red circles), 0.66 nm (black squares). The dashed lines are guides to the eye. For Fig. 4(A) and 4(B), the NC position is relative to the center of the MgO spacer, with negative numbers for NC location near the free layer and positive numbers for NC location near the hard layer [Fig. 4(C)]. An interesting feature is the W-shape in most of the $J_1$ versus NC position line cuts, which are fairly symmetric about the zero NC position (center of MgO spacer). The W-shape curves show that the coupling can be tuned in strength by changing the location of the NC within the MgO spacer. Looking at an MgO thickness of 0.70 nm and starting from the most negative position, we see $J_1$ has a similar value found in the Fe/MgO/Fe of Sample A. This is expected since the NC have merged with the free layer, resulting in a pure Fe/MgO/Fe system. As the NC move away from the free layer interface towards the center of the MgO spacer, we see that the AF coupling becomes stronger, reaching a value of $J_1 = -0.13$ erg/cm$^2$ at a NC position of -0.28 nm. When the NC approaches the zero position in the MgO spacer (middle), the AF coupling decreases in strength, reaching a minimum $J_1 = -0.005$ erg/cm$^2$. Now, as the NC moves towards the hard layer interface, $J_1$ increases in AF strength ($J_1 = -0.10$ erg/cm$^2$ at NC position +0.25 nm) before decreasing to a value similar in Sample A at the positive end point. For MgO thickness = 0.75 nm, the same trend is observed but the $J_1$ switches from AF to FM as the NC moves to the middle of the MgO spacer showing that the sign of $J_1$ can even be switched by NC position. The tailoring of $J_1$ can be further seen in a contour plot of $J_1$ as a function of Fe NC position (x-axis) and MgO thickness (y-axis) [Fig. 4(B)]. The magnitude of $J_1$ is fairly symmetric about the central Fe NC position (vertical dashed line - Black). This can be seen by the symmetric contour lines for negative $J_1$ values (blue region), which is expected due to the symmetric W-shape trend seen in the line cuts. At higher MgO thickness (> 0.85nm), there is a slight positional asymmetry with the ferromagnetic peak off center. This might be due a growth-induced asymmetry caused by vertical diffusion.

Next, we examine the coupling in the Fe/MgO/Co system with Fe NC embedded in the MgO spacer (denoted as Sample C). Fig. 5(A) and Fig. 5(B) are the representative line cuts and contour plot, respectively, for Sample C and Fig. 5(C) is the NC position index. In Fig. 5(A), we again see the W-shape trend in $J_1$ with respect to NC position, but there is a strong asymmetry in the AF coupling strength. Examining the line cut at MgO thickness of 0.72 nm (red triangles) and starting from the negative end point where the NC are at the MgO/Co interface, we find $J_1 = -0.03$ erg/cm$^2$. As the NC move vertically toward the zero NC position, we find a maximum AF coupling of $J_1 = -0.10$ erg/cm$^2$ at a Fe NC position of -0.20 nm. As the NC continue to move, the coupling reaches a minimum AF coupling of $J_1 = -0.07$ erg/cm$^2$ at a Fe NC position of -0.08 nm. With the NC continuing to move towards the hard layer, $J_1$ reaches another AF maximum of -0.16 erg/cm$^2$ at a Fe NC position of +0.22 nm. Once the NC merge with the Fe at hard layer (positive end point), the AF coupling decreases to $J_1 = -0.05$ erg/cm$^2$. The asymmetry in AF coupling is very prominent for MgO thickness of 0.66 nm (black diamonds) where $J_1 = -0.40$ erg/cm$^2$ at NC position of +0.19 nm and $J_1 = -0.21$ erg/cm$^2$ at NC position of -0.20 nm. Figure 5(B) is the resulting $J_1$ contour plot for Sample C. At high MgO thickness, we do not observe a prominent ferromagnetic peak, as was observed for Sample B [Fig. 4(B)]. Further, the asymmetry in the AF coupling can be clearly seen from differences in $J_1$ (color intensity) and the asymmetry in shape of the contour lines about the zero position line (vertical dashed line).

### IV. ANALYSIS AND DISCUSSION

#### A. Model for coupling for FM/MgO/FM with NC

To gain an insight into the origin of the features seen in Samples B and C, we develop a model for the coupling based on additional experimental observations. First, we establish that the magnetic property of the Fe NC is the most important (as opposed to the electronic property) for the effects seen in the coupling. This is supported by measurements of samples with non-magnetic NC (Ag and Al), where it is found that the W-shape feature in the line cuts of $J_1$ vs. NC position are lost. Thus, we assume the simplest magnetic coupling, which is just pair-wise bilinear coupling among the magnetic elements: the coupling between the free and hard layer ($J_{Hard\text{-}Free}$), the coupling between the free layer and NC ($J_{Free\text{-}NC}$), and the coupling between the hard layer and NC ($J_{Hard\text{-}NC}$) as shown in Fig. 6(A). Second, coupling in samples with Fe NC showed little temperature dependence and no presence of biquadratic coupling.[28] Therefore, we ignore the effect of thermal fluctuations and assume that the system is at a minimum energy. Based on these assumptions, the total bilinear coupling between the hard and free layers is (see appendix)

$$J_1(t_1, t_2) = J_{Hard-Free}(t) + \frac{|J_{Free-NC}(t_1) + J_{Hard-NC}(t_2)| - |J_{Free-NC}(t_1) - J_{Hard-NC}(t_2)|}{2} \quad (1)$$

where $t$ is the total MgO spacer thickness, $t_1$ ($t_2$) is the MgO thickness between the free (hard) layer and NC ($t = t_1 + t_2$). The first term represents the direct coupling, while the second term represents the effect of coupling to the Fe NC.

To see if this model has the same qualitative features as the data, namely the symmetric W-shape for Fe/MgO/Fe and the asymmetric W-shape for Fe/MgO/Co, we assume a functional form for the NC-FM layer coupling that is similar to $J_{Hard\text{-}Free}(t)$. Figure 6(B) shows the assumed form of the coupling, $j(t)$, as a function of thickness, which is based on a double exponential fit of the Fe/MgO/Fe data in Sample A and the fact that the coupling is ferromagnetic at low thickness. Including a strength scaling factor, $A$, we have

$$J_{Free-NC}(t_1) = A_{Free-NC} j(t_1) \quad (2)$$

$$J_{Hard-NC}(t_1) = A_{Hard-NC} j(t_2) \quad (3)$$

In Fig. 6(C), $J_1$ is plotted as a function of NC position with $t = 0.7$ nm and coupling strengths for FM layer to NC are set to be equal for the free and hard layer ($A_{Hard-NC} = A_{Free-NC} = 0.08$). Comparing the simulation [Fig. 6(C)] to the line cut at MgO thickness of 0.7 nm in Sample B [Fig. 6(D)], the model qualitatively reproduces the characteristic W-shape trends observed for the Sample B line cut. The model provides an intuitive explanation for the shift toward ferromagnetic coupling when the magnetic NC is near the center of the MgO spacer. Because both the free and hard layers couple antiferromagnetically to the NC, the cumulative effect is that the magnetizations of the two layers want to be parallel to each other. To further test the model, we calculate a $J_1$ contour plot using Eq. (1)-(3) [Fig. 6(E)] and compare to the $J_1$ contour plot of Sample B [Fig. 4(B)], showing again that the pair-wise coupling model can capture the features seen in our samples.

Next, we try to produce the asymmetry in the line cuts that were seen in the Sample C [Fig. 5(A)]. Figure 6(F) shows a simulated line cut for highly asymmetric coupling strength using $t = 0.7$ nm, $A_{Hard-NC} = 0.08$ and $A_{Free-NC} = 0.28$. This shows a strong asymmetry in the line cut that is similar to the data of Sample C. Because the AF coupling is stronger at the positive NC position in both the data and simulation, it implies that the coupling between the Fe NC and Co (free layer) is much stronger than the coupling between the Fe NC and Fe (hard layer). Therefore, by assuming a reasonable functional form for $J_{Free-NC}$ and $J_{Hard-NC}$ (equations 2 and 3), the model defined by equation 1 is able to capture the main features of the experimental data.

### B. Coupling to Fe nanoclusters

We now turn our attention to using equation 1 to quantitatively determine the values for $J_{Free-NC}$ and $J_{Hard-NC}$. To do this, we no longer assume the functional forms of equations 2 and 3. The only assumption we make is that when the NC is very close to a FM layer, its coupling will be very strongly ferromagnetic so their magnetizations will be aligned. For the case when the NC is very close the hard layer, then equation 1 reduces to

$$J_1 \approx J_{Hard-Free} + J_{Free-NC} \tag{4}$$

For the case when the NC is very close the free layer, then equation 1 reduces to

$$J_1 \approx J_{Hard-Free} + J_{Hard-NC} \tag{5}$$

In either case, the total coupling is the sum of the coupling between the hard and free layer ($J_{Hard-Free}$) and the coupling between the NC and the distant FM layer.

In order to isolate the coupling between the NC and the FM layer, it is therefore necessary to determine the value of $J_{Hard-Free}$. Fortunately, this is possible for the case when the Fe NC is near a Fe layer. When the Fe NC is located directly at the Fe layer, the NC merges with the Fe layer and the NC ceases to exist as a separate entity. In this limit, the total coupling is just given by $J_{Hard-Free}$. Thus, the value of $J_1$ at the end point of the line cut is $J_{Hard-Free}$, and the variation of $J_1$ away from the end point is equal to the coupling between the NC and the distant FM layer.

For Sample C, this applies to the case when the NC is close to the Fe (hard) layer, yielding values for the coupling between the NC and Co layer. Figure 7 shows a line cut at total MgO thickness of $t = 0.66$ nm. The value of $J_1$ at a standardized distance of 0.1 nm between the NC and Fe layer yields a value of $J_{Co-NC} = -0.26$ erg/cm$^2$ as illustrated in the figure. This procedure is repeated for each line cut in Figure 5(A) to obtain the values of $J_{Co-NC}$ as a function of MgO spacing and the results are plotted in Figure 8(A) (blue circles). For Sample B, this procedure is performed for each Fe layer, yielding two data sets for the coupling between the NC and Fe layer ($J_{Fe-NC}$) as a function of distance. Figure 8(A) shows the results for the Fe(hard)-NC coupling (open red circles) and the Fe(free)-NC coupling (open red squares). For comparison with layer-to-layer coupling, in Figure 8(B) we plot the coupling in Fe/MgO/Fe and Fe/MgO/Co (without NC) over the matching MgO thickness range.

Comparing Fig. 8(A) and 8(B), we see that the $J_{Co-NC}$ is always more AF than $J_{Fe-NC}$, unlike the trend seen for $J_{Fe-Co}$ and $J_{Fe-Fe}$. For the coupling of two thin films [Fig. 8(B)], we see that $J_{Fe-Fe}$ is slightly more AF than $J_{Co-Fe}$ for MgO thicknesses above 0.65 nm, while the AF coupling of $J_{Co-Fe}$ is clearly stronger than $J_{Fe-Fe}$ for MgO thickness below 0.65 nm. At MgO thickness of ~0.56 nm, $-J_{Co-Fe} \approx 0.37$ erg/cm$^2$ and $-J_{Fe-Fe} \approx 0.31$ erg/cm$^2$, which has a difference of 0.06 erg/cm$^2$, or ~20%. For the coupling between thin film to NC [Fig. 8(A)], $J_{Co-NC}$ always has stronger AF coupling than $J_{Fe-NC}$ and at MgO thickness of 0.56 nm, $-J_{Co-NC}$ = 0.26 erg/cm$^2$ and $-J_{Fe-NC}$ = 0.10 erg/cm$^2$, which has a difference of 0.16 erg/cm$^2$, or ~160%. This is much larger than the difference between $J_{Co-Fe}$ and $J_{Fe-Fe}$, both in terms or percentage difference and in absolute magnitude.

Qualitatively, the magnitude of coupling and coupling differences should scale with the area of the FM/MgO interface, which is smaller for a layer of NC than a continuous film. Although the magnitude of the coupling does decrease in the NC systems, the decrease is not nearly as much as one would expect based on the reduced area of the ¼ ML Fe NC. Interestingly, we find that the difference in the coupling between Fe/MgO/Fe and Fe/MgO/Co systems is amplified when the Fe is reduced from a thin film to a NC layer. Further studies will be needed to understand the microscopic origin of this nano-scaling effect.

## V. CONCLUSION

We measured the interlayer exchange coupling across the Fe/MgO/Fe and Fe/MgO/Co systems with and without embedded Fe nanoclusters. First, we find that changing the material composition of the free layer from Fe to Co/Fe enhances the coupling across MgO. Next, by embedding Fe NC at different positions within the MgO spacer in both Fe/MgO/Fe and Fe/MgO/Co systems, we can tailor the strength and sign of $J_1$. Through developing a pair-wise coupling model, we show that the observed effects are due to the magnetic coupling between the FM layers and NC. Lastly, we compare differences in coupling observed in the thin-film/NC systems (Co/MgO/NC and Fe/MgO/NC) to the analogous thin-film systems (Fe/MgO/Co and Fe/MgO/Fe) and find that the coupling difference is greater in the NC systems, providing evidence for enhanced material dependence in $J_1$ due to nano-scaling effects.

## VI. ACKNOWLEDGEMENTS

We would like to acknowledge K. M. McCreary, K. Pi, Y. F. Chiang, X. Tan, and H. W. K. Tom for their support and stimulating discussions throughout the project. This work has been supported by NSF (DMR-0706681) and CNN (DMEA-H94003-09-2-0904).

## APPENDIX

For the FM/MgO/FM system with FM nanoclusters embedded in the spacer, the energy equation of the system is,

$$E = -J_{Hard-Free}\cos(\theta_H - \theta_F) - J_{Free-NC}\cos(\theta_F - \theta_{NC}) - J_{Hard-NC}\cos(\theta_H - \theta_{NC}) \\ + K_H t_H \cos^2(\theta_H)\sin^2(\theta_H) + K_F t_F \cos^2(\theta_F)\sin^2(\theta_F) + K_{NC} t_{NC} \cos^2(\theta_{NC})\sin^2(\theta_{NC})$$ (A1)

Where $\theta_H$, $\theta_F$, and $\theta_{NC}$ are the in-plane magnetization angle relative to the applied field direction for the hard layer, free layer and NC, respectively, and $K_H$ ($t_H$), $K_F$ ($t_F$), and $K_{NC}$ ($t_{NC}$) are the corresponding anisotropy (thickness), respectively. The anisotropy parameters are assumed to be positive, which is the case for Fe and bcc Co. Assuming that the magnetization of the hard layer is fixed ($\theta_H = 0°$) leads to

$$E = -J_{Hard-Free}\cos(\theta_F) - J_{Free-NC}\cos(\theta_F - \theta_{NC}) - J_{Hard-NC}\cos(\theta_{NC}) \\ + K_F t_F \cos^2(\theta_F)\sin^2(\theta_F) + K_{NC} t_{NC} \cos^2(\theta_{NC})\sin^2(\theta_{NC})$$ (A2)

The total bilinear coupling is given by,

$$J_1 = \frac{E(\theta_F = 180°) - E(\theta_F = 0°)}{2}$$ (A3)

where

$$E(\theta_F = 0°) = -J_{Hard-Free} - J_{Free-NC}\cos(\theta_{NC}) - J_{Hard-NC}\cos(\theta_{NC}) + K_{NC} t_{NC}\cos^2(\theta_{NC})\sin^2(\theta_{NC})$$ (A4)

$$E(\theta_F = 180°) = J_{Hard-Free} - J_{Free-NC}\cos(\theta_{NC}) + J_{Hard-NC}\cos(\theta_{NC}) + K_{NC} t_{NC}\cos^2(\theta_{NC})\sin^2(\theta_{NC})$$ (A5)

Minimizing equations (A4) and (A5), $dE/d\theta_{NC} = 0$, we find that $\theta_{NC} = 0°$ or $180°$ and the lower value of $E$ yields

$$E(\theta_F = 0°) = -J_{Hard-Free} - |J_{Free-NC} + J_{Hard-NC}|$$ (A6)

$$E(\theta_F = 180°) = J_{Hard-Free} - |J_{Free-NC} - J_{Hard-NC}|$$ (A7)

Inserting Eqs. (A6) and (A7) into Eq. (A3), we find

$$J_1(t_1, t_2) = J_{Hard-Free}(t) + \frac{|J_{Free-NC}(t_1) + J_{Hard-NC}(t_2)| - |J_{Free-NC}(t_1) - J_{Hard-NC}(t_2)|}{2}$$ (A8)

Figure Captions

FIG. 1. (A) Complete layer structure for the Fe/MgO/Fe system, (B) Complete layer structure for the Fe/MgO/Co system, [(C)-(E)] RHEED patterns for the MgO buffer layer, Fe (15 nm) free layer after annealing, and Co(4 ML)/Fe(15 nm), respectively, (F) Typical RHEED intensity oscillations for Co growth on Fe (15nm), (G) Representative major hysteresis loop (dashed line) and the corresponding minor hysteresis loop (solid line) for Fe/MgO/Fe with MgO thickness = 0.67 nm.

FIG. 2. (A) Geometry of Sample A, with Fe/MgO/Co (left side) and Fe/MgO/Fe (right side) grown on the same sample and with a wedged MgO spacer, (B) Bilinear coupling $J_1$ as a function of MgO thickness for Fe/MgO/Fe (white squares) and Fe/MgO/Co (black circles).

FIG. 3. Schematic for the double-wedge MgO spacer used for Sample B and Sample C. Moving from point A to point C, the total MgO thickness is constant while the position of the Fe nanoclusters (NC) changes. Moving from point B to point D, the total MgO thickness changes.

FIG. 4. Bilinear coupling in Sample B: Fe/MgO/Fe with Fe NC. (A) $J_1$ as a function of the NC position at MgO thicknesses of 1.04 nm (blue circles), 0.85 nm (green diamonds), 0.75 nm (orange triangles), 0.70 nm (red circles), and 0.66 nm (black squares), (B) Contour/color plot of $J_1$ with red for FM regions, blue for AF regions, and thick black line for the $J_1 = 0$ erg/cm$^2$ contour. Green, orange and red dashed lines correspond to line cuts at MgO thickness of 0.85 nm, 0.75 nm and 0.70 nm, respectively, (C) The NC position value index for Sample B.

FIG. 5. Bilinear coupling in Sample C: Fe/MgO/Co with Fe NC. (A) $J_1$ as a function of the NC position at MgO thicknesses of 1.05 nm (grey squares), 0.85 nm (green circles), 0.79 nm (orange triangles), 0.75 nm (blue diamonds), 0.72 nm (red triangles), and 0.66 nm (black squares). (B) Contour/color plot of $J_1$ with red for FM regions, blue for AF regions, and thick black line for the $J_1 = 0$ erg/cm$^2$ contour. Green, orange, and red dashed lines correspond to line cuts at MgO thickness of 0.85 nm, 0.79 nm and 0.72 nm, respectively, (C) The NC position value index for Sample C.

FIG. 6. (A) Schematic of the pair-wise coupling model, (B) Plot of the $j(t)$ function used in the simulation. (C) Simulated line cut for MgO thickness of 0.70 nm, (D) line cut data for Sample B at $t = 0.70$ nm. (E) Simulated $J_1$ contour plot, (F) Simulated line cut for $A_{Free-NC} = 0.28$ and $A_{Hard-NC} = 0.08$.

FIG. 7. The method for determining the values of $J_{Hard-Free}$ (horizontal line) and $J_{FM-NC}$ (vertical line), the coupling between the Fe NC and FM layer in line cut data for Sample C at $t = 0.66$ nm. The dashed line is a guide for the eye.

FIG. 8. (A) $-J_{FM-NC}$ as a function of MgO thickness. Blue circles are for Co/MgO/NC, open red circles are for Fe(hard)/MgO/NC, and open red squares are for Fe(free)/MgO/NC. (B) Coupling observed in Fe/MgO/Co (blue circles) and Fe/MgO/Fe (red triangles) in Sample A.

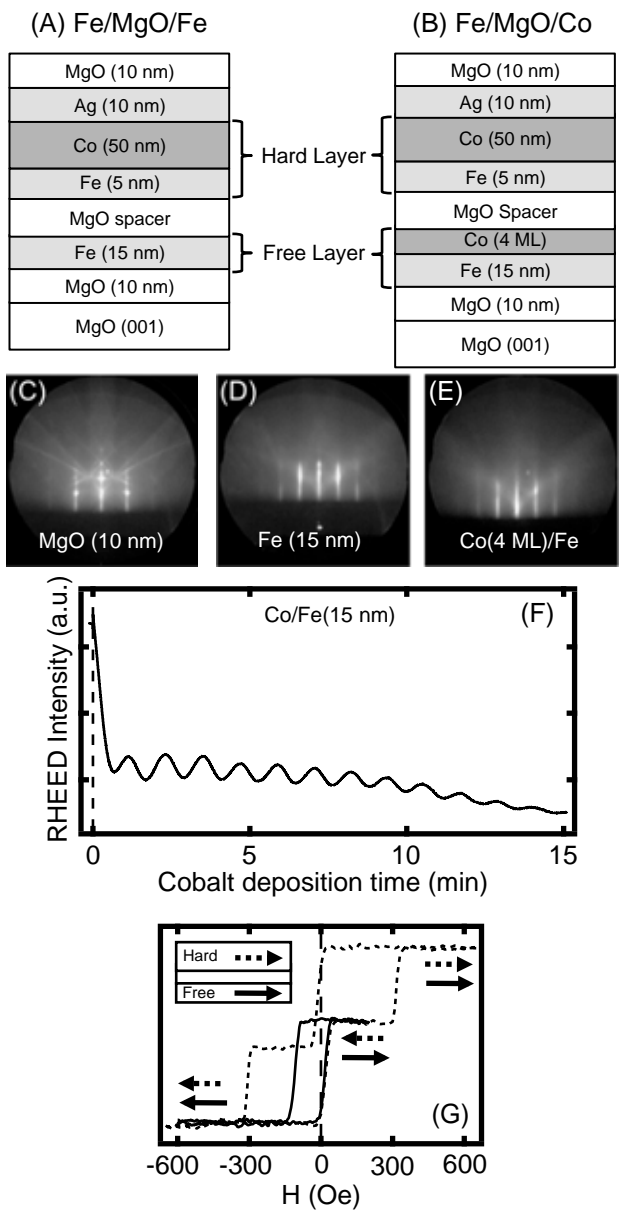

Figure 1

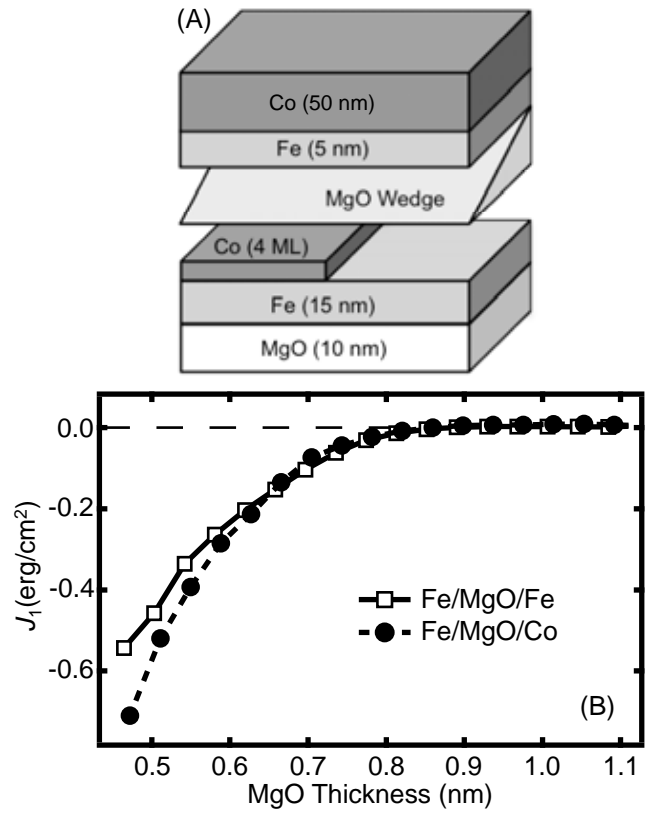

Figure 2

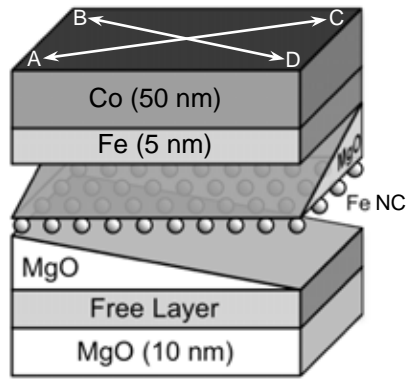

Figure 3

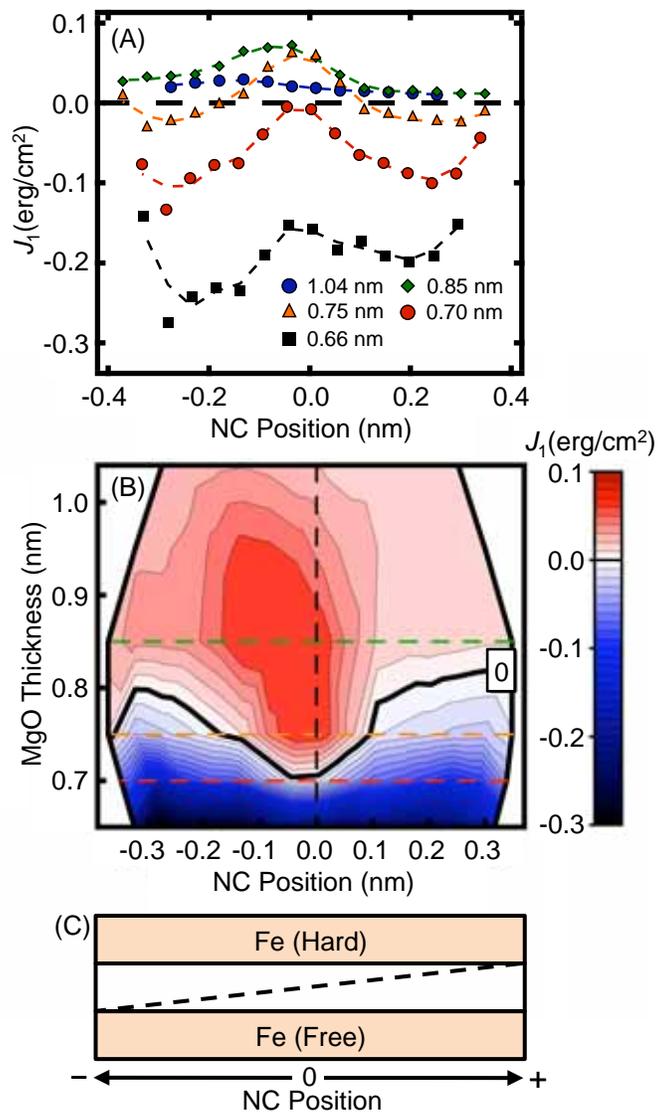

Figure 4

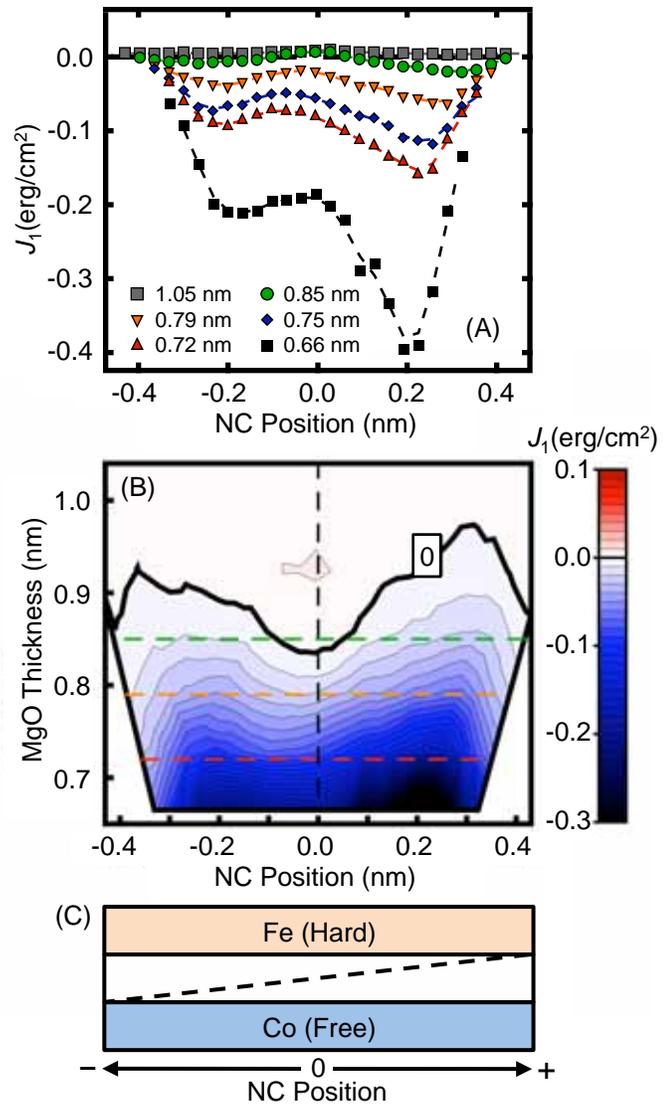

Figure 5

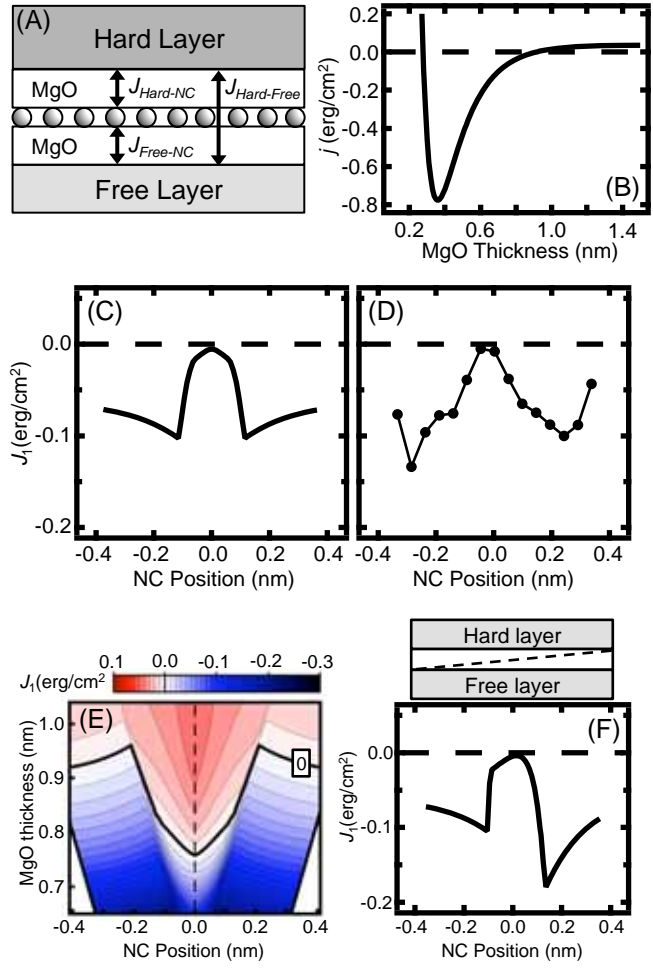

Figure 6

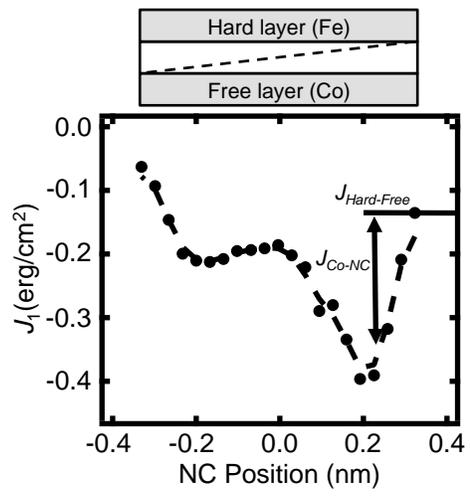

Figure 7

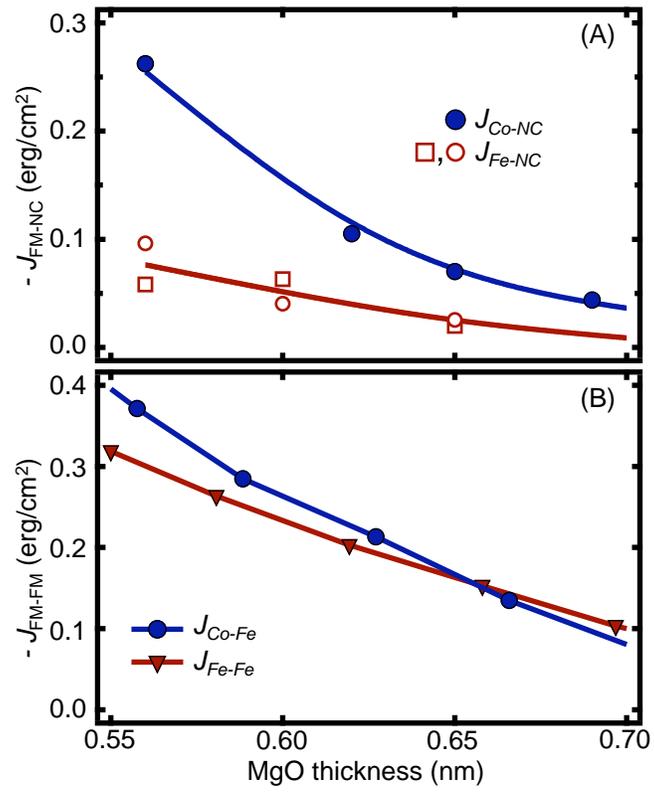

Figure 8